\documentclass[fleqn,10pt]{wlscirep}
\usepackage[utf8]{inputenc}
\usepackage[T1]{fontenc}
\usepackage{gensymb}
\usepackage{multirow}
\usepackage{graphicx}
\usepackage{subfigure}
\usepackage{caption}
\usepackage{epstopdf}
\usepackage{epsfig}
\usepackage[ruled,vlined]{algorithm2e}
\usepackage{colortbl}
\definecolor{mygray}{RGB}{200, 200, 200}
\title{A Deep Convolutional Neural Network-Based Novel Class Balancing for Imbalance Data Segmentation}

\author[1]{Atifa Kalsoom}
\author[1]{M.A. Iftikhar}
\author[2*]{Amjad Ali}
\author[2*]{Zubair Shah}
\author[3]{Shidin Balakrishnan}
\author[4]{Hazrat Ali}
\affil[1]{Department of Computer Science, COMSATS University Islambad, Lahore Campus, Pakistan.}
\affil[2]{College of Science and Engineering, Hamad Bin Khalifa University, Doha, Qatar.}
\affil[3]{Hamad Medical Corporation, Qatar.}
\affil[4]{University of Stirling, UK.}
\affil[*]{amjad.khu@gmail.com; zshah@hbku.edu.qa}
\affil[+]{Atifa Kalsoom and Amjad Ali contributed equally to this work}

\begin{abstract}
Retinal fundus images provide valuable insights into the human eye's interior structure and crucial features, such as blood vessels, optic disk, macula, and fovea. However, accurate segmentation of retinal blood vessels can be challenging due to imbalanced data distribution and varying vessel thickness. In this paper, we propose BLCB-CNN, a novel pipeline based on deep learning and bi-level class balancing scheme to achieve vessel segmentation in retinal fundus images. The BLCB-CNN scheme uses a Convolutional Neural Network (CNN) architecture and an empirical approach to balance the distribution of pixels across vessel and non-vessel classes and within thin and thick vessels. Level-I is used for vessel/non-vessel balancing and Level-II is used for thick/thin vessel balancing. Additionally, pre-processing of the input retinal fundus image is performed by Global Contrast Normalization (GCN), Contrast Limited Adaptive Histogram Equalization (CLAHE), and gamma corrections to increase intensity uniformity as well as to enhance the contrast between vessels and background pixels. The resulting balanced dataset is used for classification-based segmentation of the retinal vascular tree. We evaluate the proposed scheme on standard retinal fundus images and achieve superior performance measures, including an area under the ROC curve of 98.23\%, Accuracy of 96.22\%, Sensitivity of 81.57\%, and Specificity of 97.65\%. We also demonstrate the method's efficacy through external cross-validation on STARE images, confirming its generalization ability.

\end{abstract}
\begin{document}

\flushbottom
\maketitle

\thispagestyle{empty}

\section*{Introduction}
Human visual system is a complex network that has been studied for centuries, yet there is still much to learn about its workings. One crucial component of the visual system is the retina, a thin layer of photosensitive tissue located at the back of the eye~\cite{1}. The retina converts incoming light into neural signals that are then processed by the brain to create visual images. As the retina plays such a crucial role in vision, its health is of utmost importance. Damage to the retina can result in permanent blindness, as well as various vision impairments, such as age-related macular degeneration, cataract, diabetic retinopathy, glaucoma, and amblyopia~\cite{2}. Therefore, the precise and timely diagnosis and analysis of retinal images is highly desirable to mitigate the aggravating effects of retinal pathologies and improve disease prognosis ~\cite{3, 4, 400}. The retinal fundus images are widely utilized non-invasively by medical professionals for analyzing visual impairments \cite{5}. The blood vessels in the retinal fundus images form a tree structure with varying thicknesses of vessel branches, and are hence referred to as thick and thin vessels. Thick vessels are the basis for vessel diameter measurement, while thin vessels are considered a primary source for microaneurysm detection \cite{6}. Delineation and quantification of these blood vessels may serve as fundamental biomarkers for the detection and analysis of diabetic retinopathy and other visual impairments \cite{7, 8, 9, 10, 10-A}. However, manually segmenting these vessels accurately, especially thin vessels, is a difficult problem due to their poor contrast \cite{11}. Such manual segmentation is also time-consuming and requires much observer intervention, higher cost, and time availability, which leads to delays in the diagnosis and recommendation of preliminary treatment. An accurate automated segmentation for retinal blood vessels is highly desirable, which can tackle poor contrast and noise in the image \cite{12, 13, 14}. Accurate vessel segmentation can help clinicians plan and execute treatments more effectively, such as in radiation therapy or surgery. 
\begin{figure*}[t!]
\centering
\includegraphics[width=0.8\textwidth]{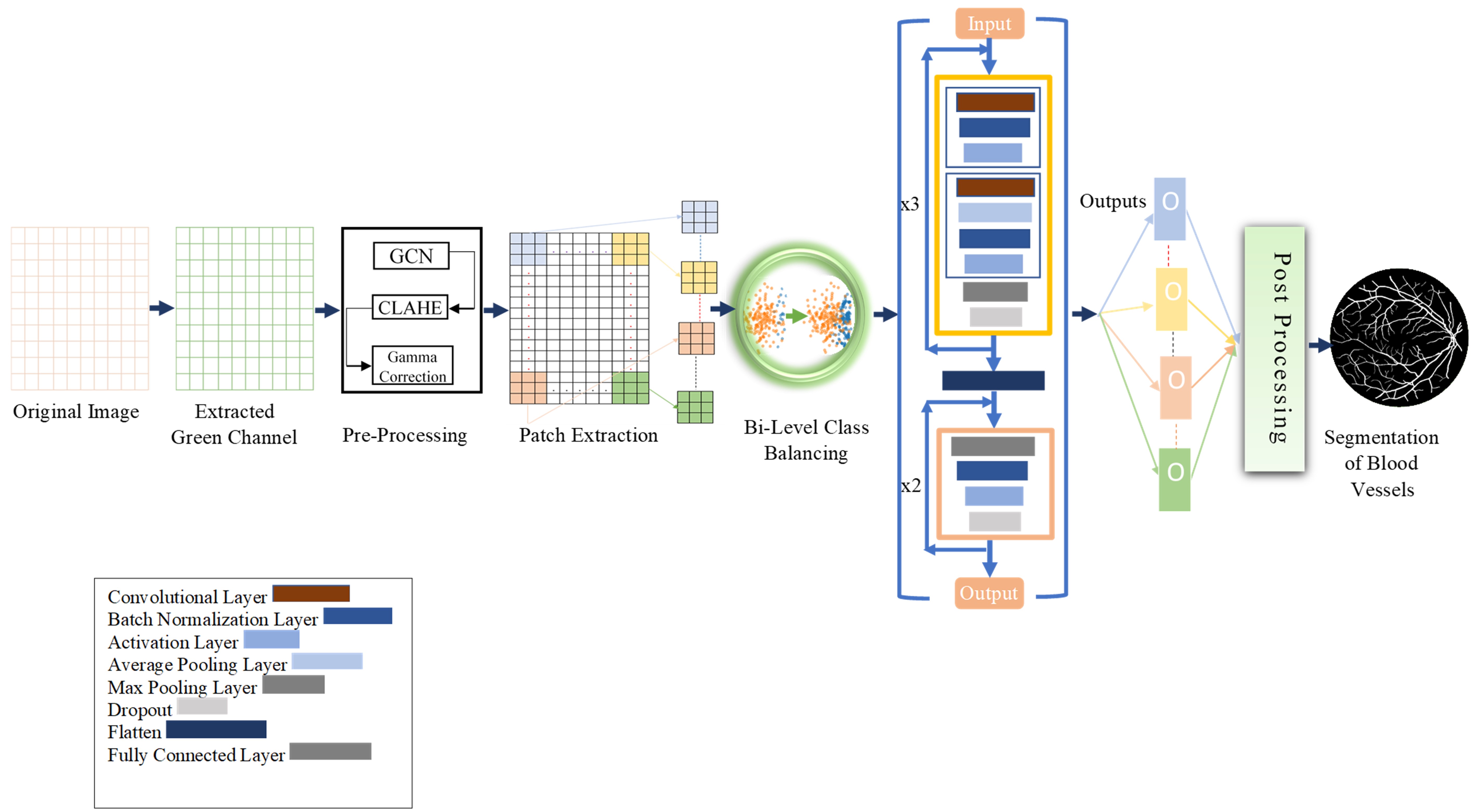}
\caption{Top-level architectural diagram of the proposed BLCB-CNN Scheme.}
\label{fig}
\end{figure*}
\par Contemporary literature supports many algorithms, techniques, and models aiming at precise and accurate segmentation of retinal blood vascular tree \cite{15}. A lot of these efforts share the common challenge of accurately segmenting the tiny blood vessels. However, deep learning-based methodologies have continuously demonstrated their ability to learn robust features for precise blood vessel segmentation automatically \cite{16, 17, 18, 19, 20, 21, 22}. For instance, Liskowski \& Krawiec \cite{23} trained and evaluated many diverse deep learning architectures combined with a structured prediction for the classification of multiple pixels simultaneously. Orlando et al. [24] segmented blood vessels by discriminatively training a fully connected Conditional Random Field (CRF) model. Model parameters are learned automatically by a structured output support vector machine (SVM). Fu et al. \cite{25}, \cite{26} modeled the blood vessel segmentation problem as a boundary detection problem and proposed a combination of a Convolutional Neural Network (CNN) and a fully connected CRF for the construction of a vessel probability map. Likewise, Dasgupta et al. \cite{27} formulated the segmentation problem as a multi-label inference problem by utilizing the combination of CNN and structured prediction. Soomro et al. \cite{28} designed a CNN architecture with a focus on boosting sensitivity coupled with some pre-processing and post-processing techniques to resolve issues of uneven illumination and removing background noise, respectively. Talha et al., \cite{28a} defined the task of segmentation as a generative task and designed a generative adversarial network (GAN) to generate synthetic retinal fundus images along with the segmented masks for the vessel tree. Hu et al. \cite{29} divided the vessel segmentation task into two phases. In the first phase, an enhanced cross-entropy loss function is used with multiscale CNN to produce the probability map. In the second phase, conditional random fields employ spatial information to obtain the final segmented binary mask. Islam \cite{30} proposed semantic segmentation of retinal blood vascular tree structure through multiscale CNN. Soomro et al. \cite{31} proposed CNN architecture in which pooling layers are swapped with strided convolutional layers. Yen et al. \cite{6} handled the segmentation task by dividing the vessel segmentation task into three steps. The first step is the segmentation of thick vessels, while the second step is the segmentation of thin vessels using the information provided by already segmented thick vessels. Finally, the two segmented vessel maps were combined using a third fusion step to obtain the final segmented image mask. 

\par \textbf{Motivation:} As the above literature studies show, deep learning techniques have been effective for the segmentation of retinal blood vascular trees from retinal fundus images. However, these techniques require a balanced distribution within their data classes for optimal efficiency. For instance, in medical image classification tasks, certain medical conditions might be more prevalent than others. A data balancing algorithm can be used to address this class imbalance by balancing the dataset, thereby improving the performance of the machine learning model. An unbalanced class distribution in the training set implies a lack of generalization of the machine learning classifiers \cite{32}. The distribution of instances across classes is biased in an unbalanced dataset, in which certain classes contain fewer instances than others. Consequently, the classification model trained with such data classifies the input samples as belonging to the more prevalent classes, though the objective would be that instances in the small classes must be properly classified for a more impactful classification \cite{33}. Retinal fundus images also have more representation of thick vessels and less representations of thin vessels \cite{34}. This unbalanced class distribution results in the suboptimal classification/segmentation of thin blood vessel pixels, resulting in a lower sensitivity. In this work, we propose deep learning-based method, which encompasses certain preprocessing techniques for contrast enhancement and post processing for noise removal. The proposed method overcomes low sensitivity issue overlooked by previous works. The main contributions of the paper are given below:

\begin{itemize}
\item We designed a 10-layered deep CNN architecture to accurately segment retinal blood vascular tree from retinal fundus images. The proposed CNN architecture is supported by a Bi-Level Class Balancing (BLCB) algorithm to deal with the small sample size issue (i.e., segmentation of thin blood vessels). The first level controls the balance between the number of vessels and non-vessels pixels (inter-class balancing). In contrast, the second level trades off for the balance between several thick and thin vessel pixels (intra-class balancing). 
\item In the proposed model, pre-processing of the input retinal fundus image is performed using Global Contrast Normalization (GCN), Contrast Limited Adaptive Histogram Equalization (CLAHE), and gamma corrections that increase intensity uniformity as well as enhance the contrast between vessels and background pixels. Thus, the proposed algorithm significantly enhances scalability and makes it easier to train it on large datasets as well as it can be easily deployed in real time. 
\item The efficacy of the proposed Bi-Level Class Balancing algorithm-based Convolutional Neural Network (BLCB-CNN) scheme is validated on a publicly available DRIVE dataset of retinal fundus images with state-of-the-art models and further evaluated on the external STARE dataset to test the model's generalization. The proposed algorithm achieves superior performance measures, including an area under the ROC curve of 98.23\%, Accuracy of 96.22\%, Sensitivity of 81.57\%, and Specificity of 97.65\%. 
\end{itemize}

\par The subsequent sections are structured as follows. Section II describes the components of the proposed BLCB-CNN approach. The model architecture, the datasets, and the experimental setup are explained sequentially in Sections III, IV, and V. Section VI presents quantitative and qualitative results for DRIVE, cross-validation for STARE, and a comprehensive comparison of state-of-the-art techniques. Finally, Section VII concludes the paper and provides further research directions.

\section*{Related Work}
Imbalance datasets make vessel segmentation difficult because the presence of blood vessels is often small in comparison to the rest of the image. To address the data imbalance challenge, several methods have been proposed in the literature, including data augmentation, transfer learning, and the use of various loss functions. In ~\cite{44}, the authors proposed a deep learning method for vessel segmentation that makes use of a multi-scale triplet CNN to capture both local and global context. To address the class imbalance, the authors also employed data augmentation techniques, such as rotation, scaling, and flipping. The authors categorize retina vessels into three types: arteries, veins, and background vessels by training a multi-scale interactive network (MIN) that considers both local and global contextual information. The MIN comprised a multi-scale feature extraction network and a multi-scale interactive network. The feature extraction network extracts features at multiple scales using convolutional neural networks (CNNs), while the interactive network integrates the features to produce vessel segmentation maps. By exploiting the concept of adaptive weighted balance loss to the conventional classification network, the authors have presented a unique solution for multi-center skin lesion classification in \cite{N1}. The authors further have extended their work on imbalanced medical image classification in \cite{N2} and introduced specificity-aware federated learning scheme by exploiting the concepts from dynamic feature fusion strategy and adaptive aggregation mechanism. Similarly, in \cite{N3}, Z. Zhou et al. introduced a dynamic class balancing method for image segmentation. The authors proposed an effective sample calculation method in the context of semantic segmentation for a highly unbalanced dataset. Further, based on the effective sample concept they proposed a dynamic weighting method. In \cite{N4}, based on the concept of CNN, the authors proposed an end-to-end CNN model. The proposed model is not based on any stages of ML and only requires one stage to detect myocardial infarction from the input signals. Further, for imbalanced data, the authors optimized the proposed deep learning model with a new loss function called focal loss. 

\par In ~\cite{45}, the authors proposed SA-UNet that trained a U-Net, a popular architecture for image segmentation tasks. However, they augment the U-Net by adding a spatial attention module, which enables the network to focus on regions of interest in the image selectively. This is achieved by learning a spatial attention map that is multiplied element-wise with the U-Net's feature maps, allowing the network to emphasize important features and suppress irrelevant ones. Overall, the paper presents a novel deep learning approach for retinal vessel segmentation that leverages a spatial attention mechanism. The approach shows promising results and could be used for the automatic diagnosis of retinal diseases. To address the class imbalance problem, the authors in ~\cite{46} proposed a new loss function called the Unified Focal Loss (UFL), which generalizes both the Dice coefficient and cross-entropy loss functions. The UFL consists of two terms: a focal term that focuses on difficult-to-segment pixels and a uniform term that balances the contributions of each class. In ~\cite{47}, the authors proposed MMDC-Net, which consists of multiple layers of dilated convolutions at different scales. The dilated convolutions enable the network to have a large receptive field, which allows it to capture features at multiple scales and process the image in a more efficient way. Additionally, the authors propose a sharpened details module that helps the network to enhance the details of the segmentation results. The advantage of the MMDC-Net is that it is designed to handle images with varying degrees of vessel visibility, which is common in retinal images. The multi-scale dilated convolutions enable the network to capture features at different scales, allowing it to segment vessels of different sizes and shapes. The sharpened details module further enhances the segmentation results by improving the visual quality of the segmented vessels. The results show that the proposed approach achieves better performance than other methods in terms of accuracy and speed.

\par In~\cite{48}, the authors proposed BSEResU-Net combining the U-Net architecture with before-activation residual connections and an attention mechanism. The before-activation residual connections enable the network to learn residual mappings before the activation function, allowing it to capture more complex features. The attention mechanism is added to the network to selectively focus on regions of interest in the image, allowing the network to emphasize important features and suppress irrelevant ones. The results show that the proposed method outperforms the other methods in terms of accuracy and speed. In this paper, we propose a systematic deep learning pipeline, which utilizes a novel two-stage class balancing algorithm (named Bi-Level Class Balancing or BLCB) for handling the class imbalance issue. This balancing algorithm explicitly handles the class imbalance at both levels mentioned earlier (i.e., vessels/background pixels and thick/thin vessels). Additionally, the proposed pipeline proposes a custom CNN architecture for improved feature learning in the context of retina fundus image segmentation. Finally, the pipeline includes effective pre-processing steps for improving fundus image quality and post-processing steps for noise removal in the final segmentation. These contributions set new benchmark performance measures for the popular DRIVE dataset for retinal fundus image segmentation (AUC: 98.23\%, Accuracy: 96.22\%).
\begin{figure*}[t!]
       \centering
       \captionsetup{justification=centering}
       \subfigure[Original image in RGB]{
       \includegraphics[width=.22\textwidth]{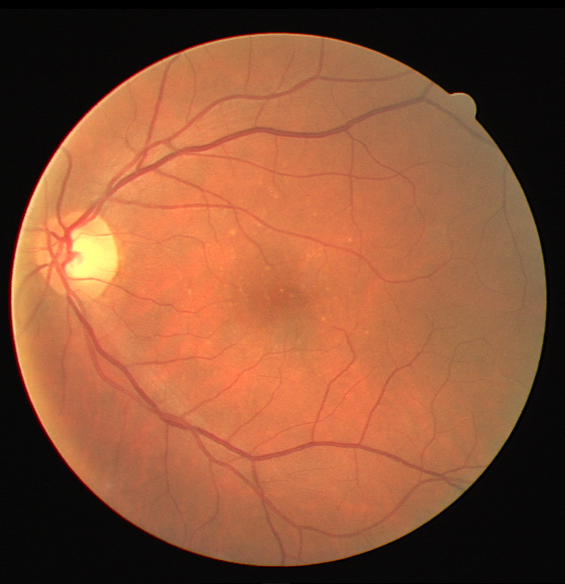}
        }
       \subfigure[Red channel]{
       \includegraphics[width=.22\textwidth]{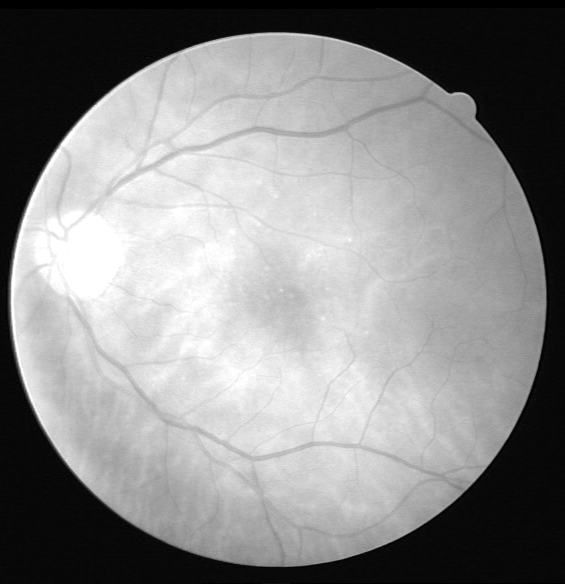}
       }
       \subfigure[Green channel]{
       \includegraphics[width=.22\textwidth]{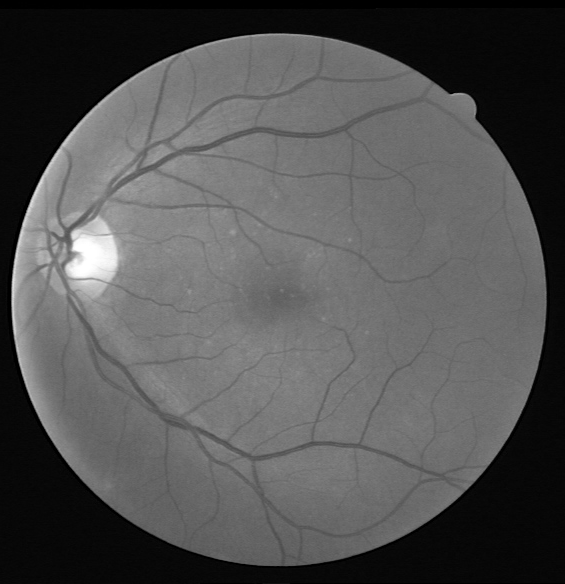}
       }
       \subfigure[Blue channel]{
       \includegraphics[width=.22\textwidth]{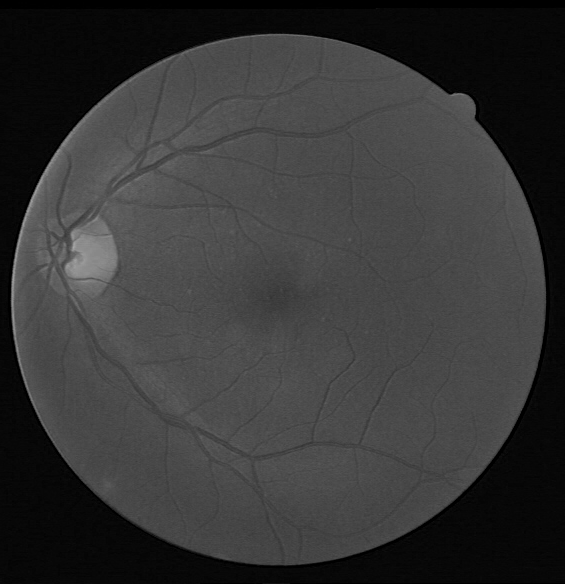}
       }
      \caption{Sample DRIVE image and its channels.}
      \label{Fig 2} 
\end{figure*}

\begin{figure*}[t!]
       \centering
       \captionsetup{justification=centering}
       \subfigure[Green channel]{
       \includegraphics[width=.22\textwidth]{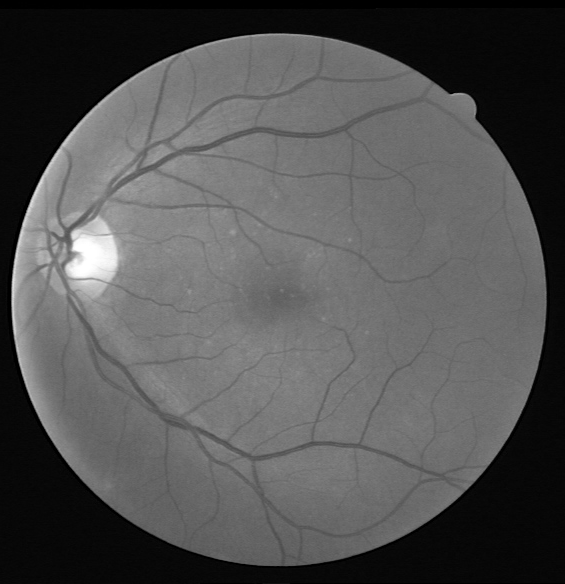}
        }
       \subfigure[Normalized green channel]{
       \includegraphics[width=.22\textwidth]{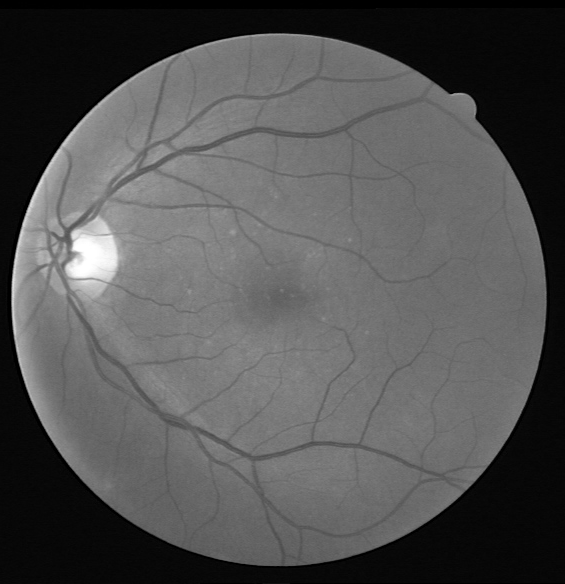}
       }
       \subfigure[Application of CLAHE on normalized green channel]{
       \includegraphics[width=.22\textwidth]{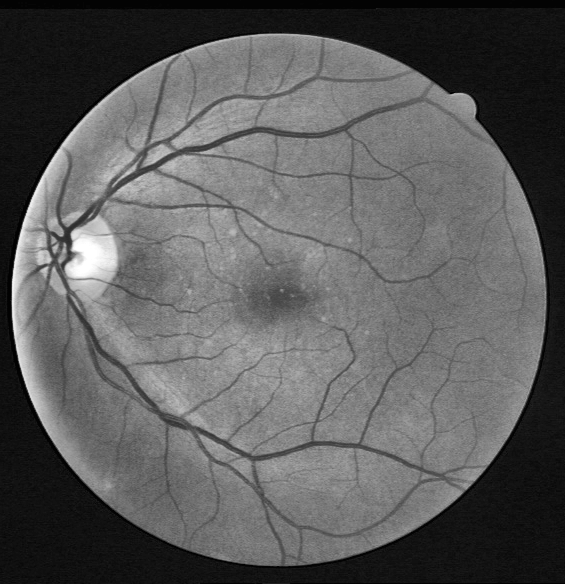}
       }
       \subfigure[Application of gamma correction]{
       \includegraphics[width=.22\textwidth]{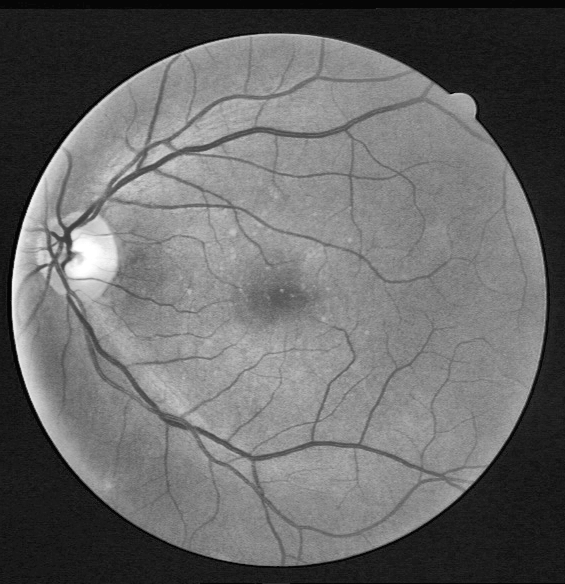}
       }
       
      \caption{Preprocessed green channel of sample DRIVE image.}
      \label{Fig 3}
\end{figure*}

\section*{Methodology}
\par The proposed BLCB-CNN methodology comprises a few components that address the challenges posed by typical retinal fundus images. Figure 1 shows the sequential workflow of these components. The proposed BLCB-CNN methodology extracts the green channel of the input retinal fundus image, which has maximum contrast among the three channels of the RGB fundus image [36]. Figure 2 depicts an original sample DRIVE image with all its separated channels, i.e., red, green, and blue. Additionally, pre-processing of the input retinal fundus image is performed using GCN \cite{36}, CLAHE, and gamma corrections to increase intensity uniformity as well as to enhance the contrast between vessels and background pixels. Figure 3 depicts the extracted green channel of a sample DRIVE image, and other images are illustrations of pre-processing steps performed on the green channel. After pre-processing, the images are input sequentially to the two major components of the proposed BLCB-CNN methodology architecture, i.e., Patch Extraction (Section II-A) and Class Imbalance Handler (Section II.B).

\begin{algorithm}
\KwIn{Preprocessed green channel of retinal fundus image $I$, Ground truth vector $Y$, and Threshold value $v$ }
\KwOut{List of patches with balanced (classes vessel vs non-vessel) $P$ }
\Begin{
  /* Read and pad image $I$ */ \\
  /* Extract patches [$i$] from image $I$ */ \\
  /* Calculate Patch Mean ($PM$) Vector from patches[$i$]: */ \\
    $~~~~ PM = \mu (patches[i])$ \\
  /* Classify the patches for length $(patches[i])$: */ \\
    
  \If{$ Y[i]~==~1$}
	 {
	    $\alpha = patches[i] $\\
	 } 
   \eIf{$ Y[i]~==~0$ \& $PM[i] \geq v $}
      {
       $\beta = patches[i] $ \\
      }
      {
       $\gamma = patches[i] $ \\
      }
   
  /* Calculate the number of patches $(y, z)$ */ \\
     $~~~~y = length(\alpha) \times \rho$	\\
     $~~~~z = length(\alpha) \times 1- \rho$ \\
  /* Select patch arrays $(pb, fb)$  Randomly */ \\
	 $~~~~pb \in \beta \sim y$ \\
      $~~~~fb \in \gamma \sim ~ z$ \\
/* Concatenate selected patches $(pb, fb)$ with vessel pixel patch array $(\alpha)$ in $P$ */ \\
 Return $P$
 } 
\caption{Level-I Balancing Algorithm } 
\end{algorithm}
\subsection*{Patch Extraction}
\par Patch extraction minimizes convolutions and boosts class instances for more reliable and effective model learning. A CNN typically requires a substantial number of images for training. Therefore, retinal fundus images divided into small sub-images (patches) are used as input to the CNN, where the patch size is a hyper-parameter. The images used in this work have high resolution (i.e., $565 \times 584$ and $605 \times 700$ for DRIVE and STARE datasets, respectively) with corresponding ground truth binary masks. Each image is padded and divided into patches of size 64x64, where the central pixel of the patch is the focal point and characterizes the class of the patch. A label based on the central pixel through one-to-one mapping from corresponding binary mask pixels is assigned to each patch. Based on the label, each patch is then classified as a vessel or non-vessel patch. The learning model learned iteratively with a batch of 64 patches and classified each patch as a vessel or non-vessel class. Batch size is a hyper-parameter as well. Both parameters (patch size and batch size) are determined by a grid search. The binary labels returned from each patch are reshaped as the original image shape to obtain the segmented binary mask of the blood vascular tree structure based on classification.

\subsection*{Class Imbalance Handler}
Artificial neural networks require a huge amount of labeled as well as balanced data for unbiased and effective training of the network. In typical retinal fundus images, non-vessel class pixels always dominate in number as compared to vessel class pixels. This imbalance in the class distribution of pixels results in lower sensitivity of many vessel segmentation methods, as the imbalance in class distribution causes biased training of the learning model towards the dominating class. To address this issue, dominating non-vessel class pixels are selected based on a random subsampling approach to match the proportion of the vessel class pixels such that both classes have a balanced distribution of pixels in the training set. The BLCB-CNN methodology includes a Bi-Level class imbalance handling algorithm for this purpose, which is described in the next section. The Bi-Level class imbalance handling algorithm works at two levels to balance the training set data.
\begin{algorithm}[t!]
\KwIn{Preprocessed green channel of retinal fundus image $I$, Ground truth vector $Y$, Threshold value $v$, ratio parameter $r$, Thick vessel pixels ground truth binary mask $V$, and Thin vessel pixels ground truth binary mask $U$ }
\KwOut{List of patches with balanced (classes vessel vs non-vessel) $P$ }
\Begin{
  /* Read and pad image $I$ */ \\
  /* Extract patches patches [$i$] from image $I$ */ \\
  /* Calculate Patch Mean ($PM$) Vector from patches[$i$]: */ \\
    $~~~~ PM = \mu (patches[i])$ \\
  /* Classify the patches for length $(patches[i])$: */ \\

   \eIf{$ U[i]~==~1$}
	 {  
	    $c\alpha = patches[i] $\\
	 } 
    {    
    \eIf{$ V[i]~==~1$}
    {  
     $n\alpha = patches[i] $ \\
    }  
    {  
    \eIf{$ Y[i]~==~0$ \& $PM[i] \geq v$ }
        {    
       $ \beta = patches[i] $ \\
        }
         {    
           $\gamma = patches[i] $ \\
          }
    }
    }  
  /* Calculate number of thick vessel patches $x$ */ \\
     $~~~~x = length(n\alpha) \times \rho$	\\
  /* Select thick vessel patches $tp$ Randomly */ \\
	 $~~~~tp \in n\alpha \sim x$ \\
  /* Concatenate selected patches */ \\
	 $~~~~ \alpha = tp + c\alpha $ \\
  /* Calculate the number of patches $(y, z)$ */ \\
     $~~~~y = length(\alpha) \times \rho$	\\
     $~~~~z = length(\alpha) \times 1- \rho$ \\
  /* Select patch arrays $(pb, fb)$  Randomly */ \\
	 $~~~~pb \in \beta \sim y$ \\
      $~~~~fb \in \gamma \sim ~ z$ \\
 /* Concatenate selected patches $(pb, fb)$ with vessel pixel patch array $~~~~(\alpha)$ in $P$ */ \\
 Return $P$
 } 
\caption{Level-II Balancing Algorithm } 
\end{algorithm}

\begin{figure*}[b!]
       \centering
       \captionsetup{justification=centering}
       \subfigure[Ground truth binary mask]{
       \includegraphics[width=.22\textwidth]{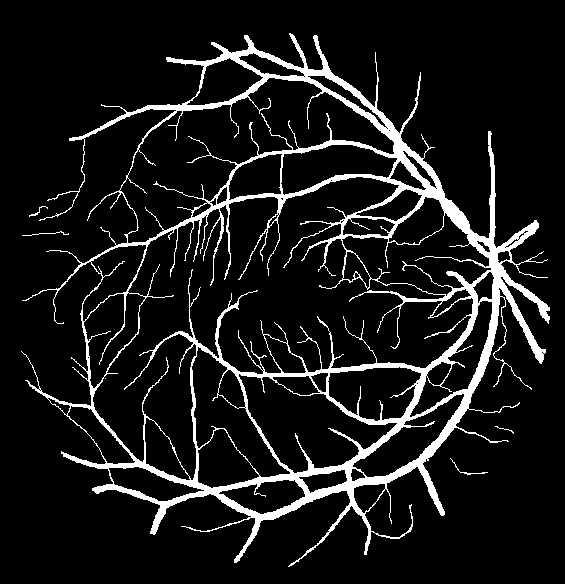}
        }
       \subfigure[Thick vessels binary mask]{
       \includegraphics[width=.22\textwidth]{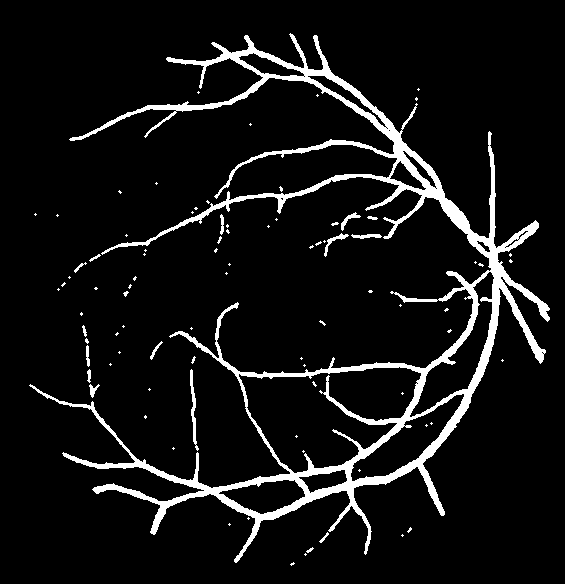}
       }
       \subfigure[Thin vessels binary mask]{
       \includegraphics[width=.22\textwidth]{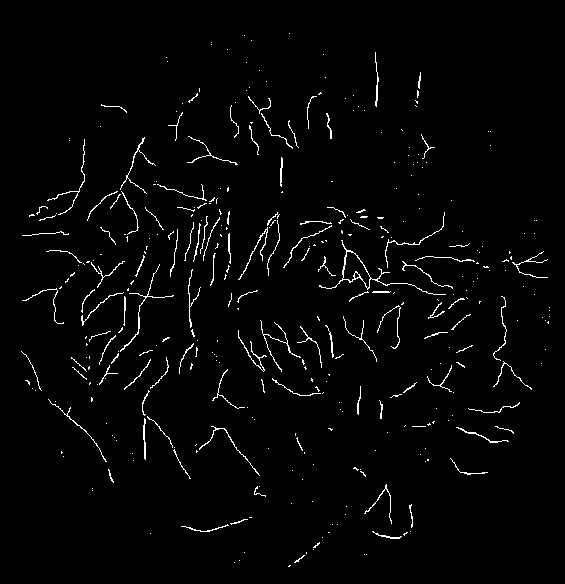}
       }
      \caption{Segregation of ground truth of sample DRIVE image.}
      \label{Fig 4}
\end{figure*}

\subsubsection*{Inter-class Balancing} 
\par This is also termed Level-I balancing, and it controls the distribution of pixels between vessel and non-vessel classes. This level resembles the non-vessel pixels to balance its distribution with the vessel class. Algorithm I shows the algorithm of Level-I balancing in detail. It takes the preprocessed green channel of the input retinal fundus image $I$, ground truth binary image $Y$, and a threshold value $v$ to distinguish between vessel and non-vessel patches. It extracts the patches, calculates the Patch Mean $PM$ vector, and classifies them in vessel pixel patch array $\alpha$, partial background pixel patch array $\beta$, and full background pixel patch array $Y$ based on $PM$. The non-vessel patches (majority class) are down-sampled by applying the random subsampling approach, which results in the same number of patches as the vessel class. To accomplish this, the algorithm determines the number of patches $(y, z)$ to be selected from partial background pixel patch array $ \beta$ and full background pixel patch array $Y$ respectively, w.r.t length of vessel pixel patch array $\alpha$. In addition, the non-vessel pixels are differentiated between simple non-vessel pixels and background pixels. The background pixels correspond to those patches, which have most of the pixels belonging to the background (completely black) part of the image. This background membership of pixels is quantified by applying an empirically determined threshold $t$ on the mean values of patches' pixels. Ideally, background pixels are the least trivial in the segmentation process; therefore, a $90: 10$ ratio $ p = 0.9 $ is maintained for simple non-vessel and background pixels, respectively. The Level-I balancing returns a list of patches $P$ with an equal distribution of vessel and non-vessel patches for training the BLCB-CNN model.

\begin{figure*}[t!]
       \centering
       \captionsetup{justification=centering}
       \subfigure[Original image]{
       \includegraphics[width=.22\textwidth]{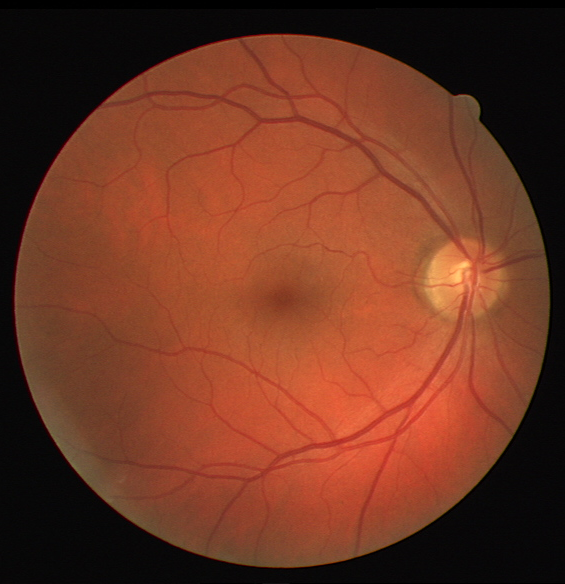}
        }
       \subfigure[Ground truth]{
       \includegraphics[width=.22\textwidth]{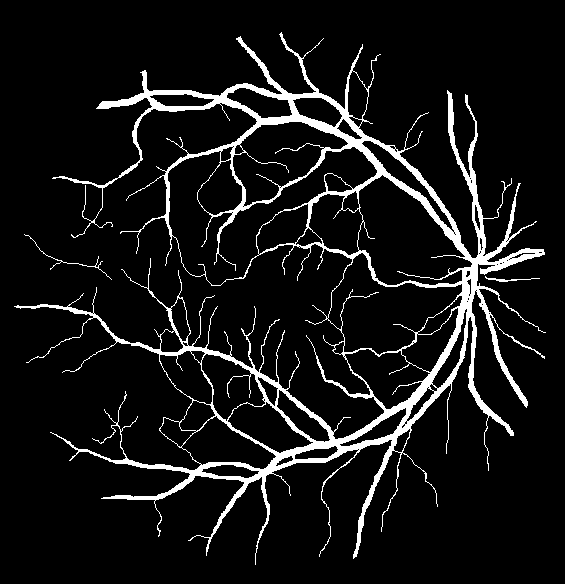}
       }
       \subfigure[Without level-I balancing]{
       \includegraphics[width=.22\textwidth]{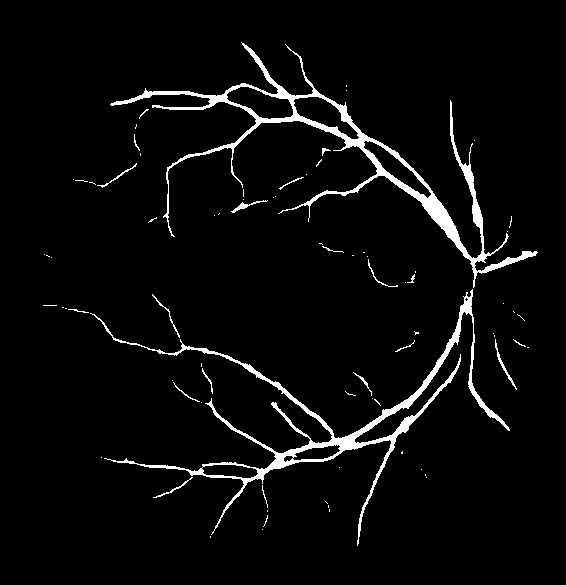}
       }
       \subfigure[With level-I balancing]{
       \includegraphics[width=.22\textwidth]{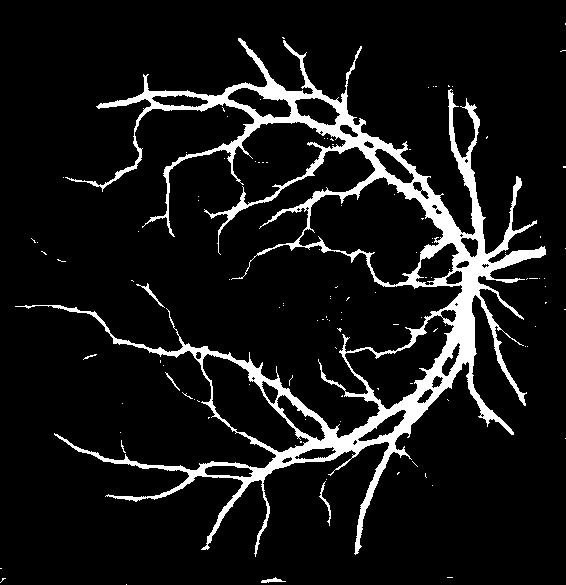}
       }
      \label{Fig 5 Row 1}
\end{figure*}

\begin{figure*}[t!]
       \centering
       \captionsetup{justification=centering}
       \subfigure[Original image]{
       \includegraphics[width=.22\textwidth]{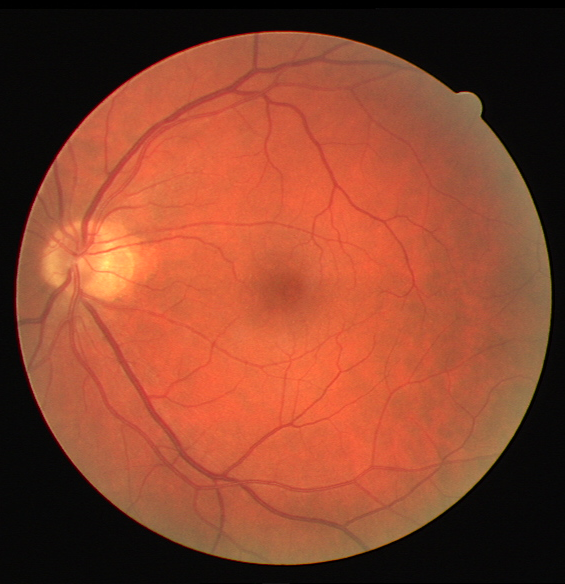}
        }
       \subfigure[Ground truth]{
       \includegraphics[width=.22\textwidth]{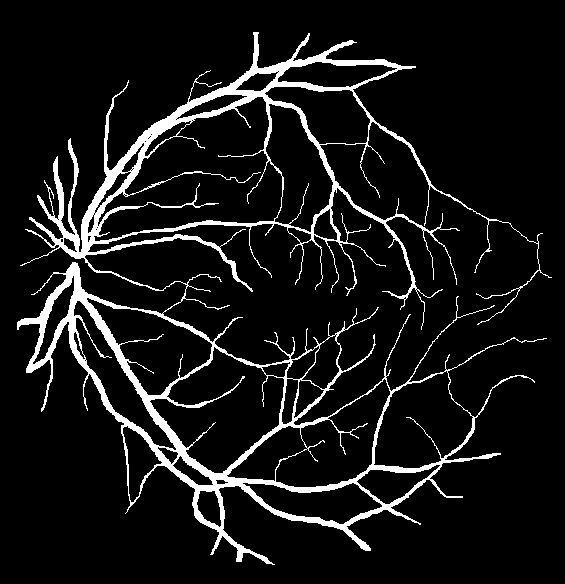}
       }
       \subfigure[Without level-I balancing]{
       \includegraphics[width=.22\textwidth]{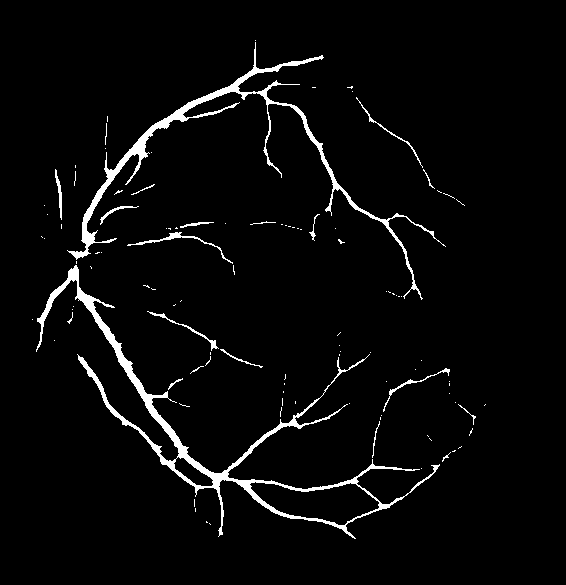}
       }
       \subfigure[With level-I balancing]{
       \includegraphics[width=.22\textwidth]{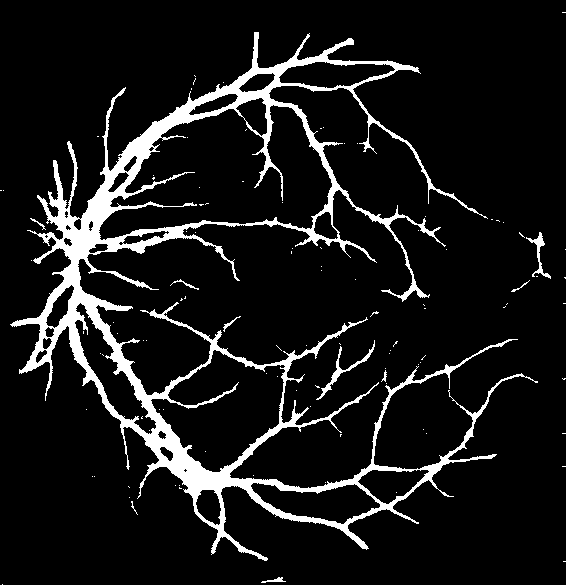}
       }
\caption{Visual results after level I balancing. The level I balancing increases the patches for vessel representation.}
      \label{Fig 5 Row 2}
\end{figure*}

\subsubsection*{Intra-class Balancing} 
\par After Level-I balancing, the network model is still biased towards the prediction of thick vessel pixels, leaving many thin vessel pixels unidentified. This is because the number of thick vessel's pixels exceeds those of thin vessel's pixels. Thus, there is a need to create a balance between the distribution of thick and thin vessel pixels in the training set, termed intra-class balancing. This can be considered another level of balancing on top of Level-I balancing, hence the term Level-II balancing. The balancing obtained from Level-I is further refined by giving an equivalent representation to thick vessel pixels and thin vessel pixels in the training set. The categorization of thick and thin vessel patches requires the segregation of ground truth binary images based on thick and thin vessel pixels. The thick vessels pixels mask is acquired through morphological opening \cite{37} operation, while the thin vessels pixels mask is obtained by subtracting the thick vessel mask from the original ground truth mask. Figure 4 shows the ground truth image of a sample DRIVE image along with the corresponding thick and thin vessel mask images.

\par The Level-II balancing algorithm is presented in Algorithm II. It takes a preprocessed green channel of retinal fundus image $I$, ground truth binary image $Y$, thick and thin vessel binary masks $U, V$, a threshold value $v$ to distinguish between vessel and non-vessel patches, and a ratio parameter $r$ to adjust the number of thick and thin vessel pixel patches. Calculate the Patch Mean vector and classify the patches into thick vessel pixel patch array $c \times \alpha$, thin vessel pixel patch array $n \times \alpha$, partial background pixel patch array $\beta$, and full background pixel patch array $\gamma$ based on $PM$. The thick vessel pixels (majority class) are down-sampled by applying the random subsampling approach, which results in the same number of pixels as the thin vessel class. The algorithm first determines the number of thick vessel pixel patches $x$ to be selected from the array of thick vessel pixel patches $n \times \alpha$ based on the adjustable ratio of thin vessel pixel patches $c \times \alpha$. Then, it concatenates selected thick vessel pixel patches $tp$ and thin vessel pixel patch array $c \times \alpha$ as vessel pixel patch array $\alpha$. Subsequently, the non-vessel patches (majority class) are down-sampled by applying the random subsampling approach, which results in the same number of patches as the vessel class. To adjust the ratio of partial and full background vessel pixel patches based on the new distribution of vessel pixels, the algorithm determines the number of patches $(y, z)$ to be selected from partial background pixel patch array $\beta$ and full background pixel patch array $\gamma$ respectively, w.r.t length of vessel pixel patch array $\alpha$. Similar to Level-I balancing, the non-vessel pixels are differentiated between simple non-vessel pixels and background pixels. Finally, the Level-II balancing returns a list of patches P with an equal distribution of patches corresponding to thick and thin vessels' pixels for training the BLCB-CNN model.

\section*{Model Architecture} 
CNN-based deep learning architectures are suitable for classification-based segmentation tasks \cite{38, 39}. Our work proposes a custom-designed CNN architecture for retinal blood vessel segmentation. The optimal deep CNN architecture in this work was determined by an exhaustive grid search on parameters, such as different numbers of convolutional layers, choice of activation and loss functions, inclusion of dropout layer, and potential learning rate values. This deep CNN architecture in the BLCB-CNN pipeline consists of three convolutional blocks, where each block consists of two sub-blocks. These blocks encode lower- and higher-level features of retinal fundus images, which are suitable for segmentation based on classification tasks. Each sub-block has a 2-D convolutional layer followed by a batch normalization layer and a ReLU activation layer. The sub-blocks proceeded by a max pooling layer and a dropout layer (dropout rate = 25\%) to avoid over-fitting. Every proceeding convolutional block doubled the filters used in the previous convolutional block. After the extraction of features through convolutional blocks, these features are passed to a flattened layer for converting them to a 1-D feature vector. This 1-D feature vector was directed to a series of fully connected layers, each of which is bundled with a batch normalization, an activation, and a dropout layer (dropout rate = 25\%). The last layer of the model is the output layer, which classifies each pixel of the image as a vessel/non-vessel pixel. These outputs for each pixel of the input image are reshaped to the dimensions of the original image and further proceeded by morphological erosion \cite{40} as a post-processing step to remove noisy/extraneous pixels. 

\begin{figure*}[t!]
       \centering
       \captionsetup{justification=centering}
       \subfigure[Original image]{
       \includegraphics[width=.22\textwidth]{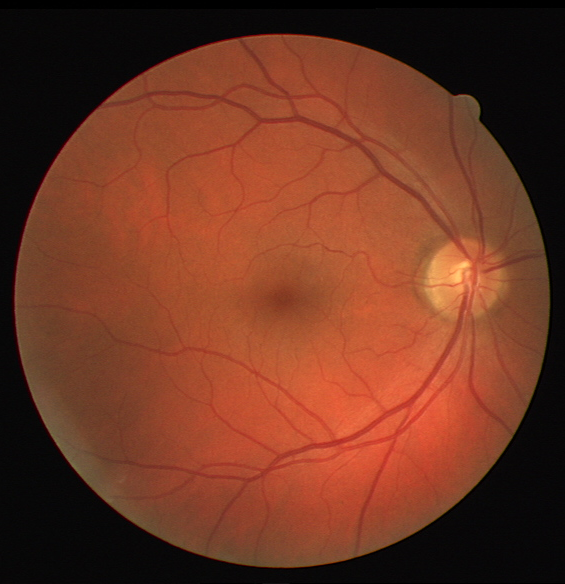}
        }
       \subfigure[Ground truth]{
       \includegraphics[width=.22\textwidth]{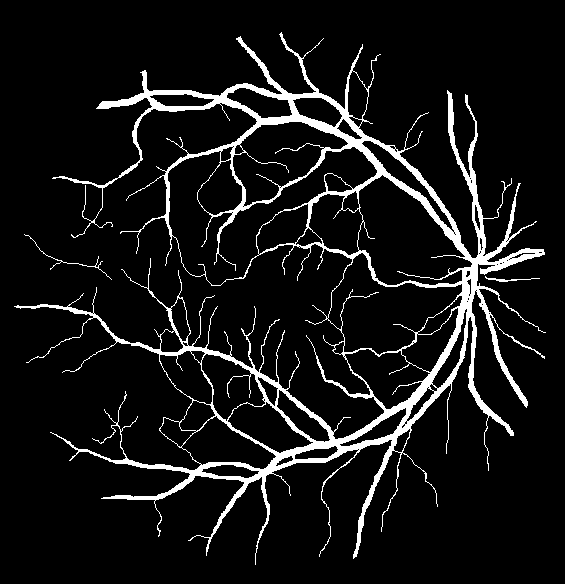}
       }
       \subfigure[Level-I balancing]{
       \includegraphics[width=.22\textwidth]{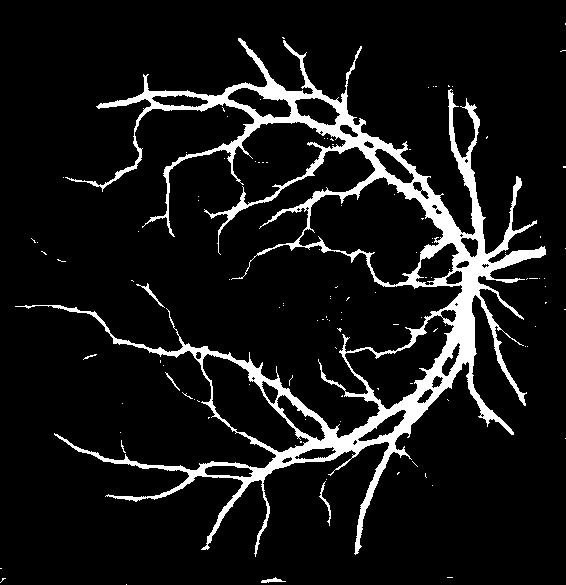}
       }
       \subfigure[Level-II balancing]{
       \includegraphics[width=.22\textwidth]{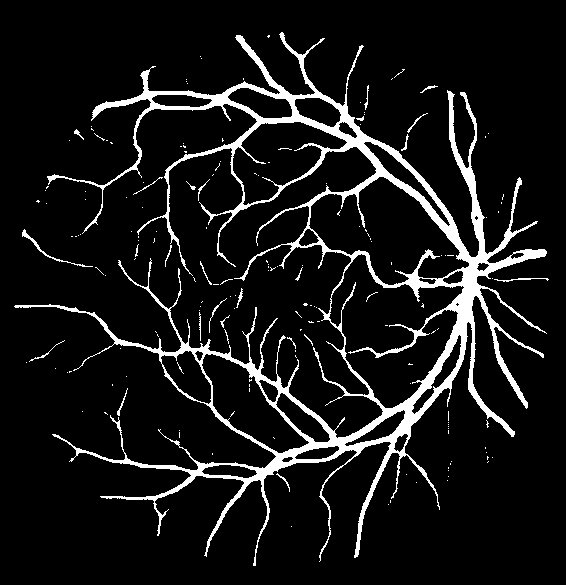}
       }
      \label{Fig 6 Row 1}
\end{figure*}

\begin{figure*}[t!]
       \centering
       \captionsetup{justification=centering}
       \subfigure[Original image]{
       \includegraphics[width=.22\textwidth]{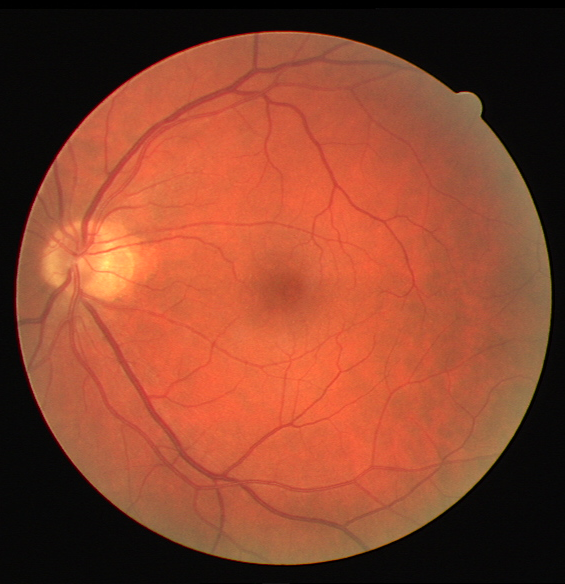}
        }
       \subfigure[Ground truth]{
       \includegraphics[width=.22\textwidth]{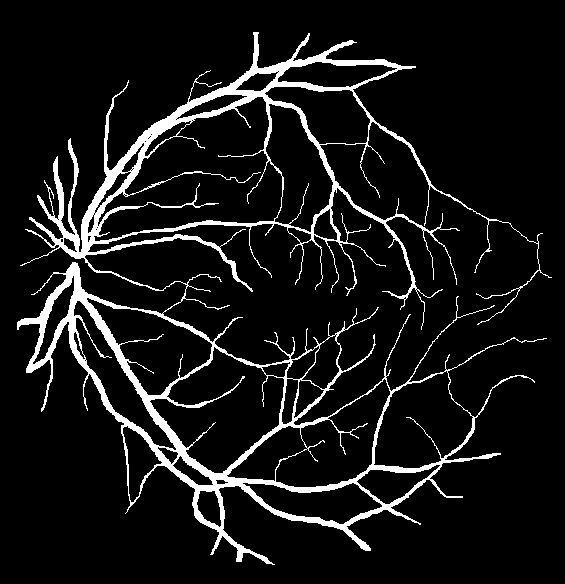}
       }
       \subfigure[Level-I balancing]{
       \includegraphics[width=.22\textwidth]{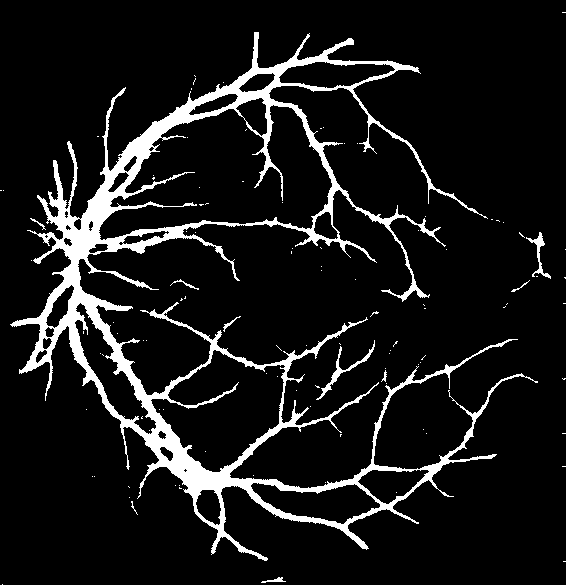}
       }
       \subfigure[Level-II balancing]{
       \includegraphics[width=.22\textwidth]{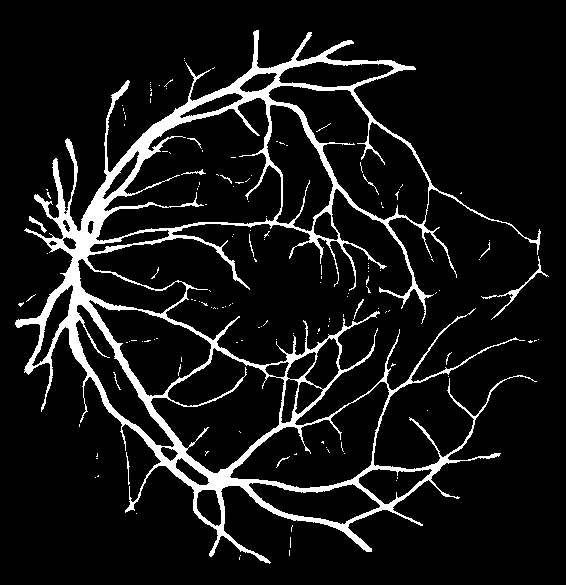}
       }
      \caption{Visual results after level-II balancing.}
      \label{Fig 6 Row 2}
\end{figure*}

\begin{table*}
\centering
\caption{Performance analysis of level-I and level-II balancing.}
\begin{tabular}{|p{1.8cm}|p{0.8cm}|p{0.8cm}|p{0.8cm}|p{0.8cm}|p{0.8cm}| p{0.9cm}|p{0.9cm}|p{0.9cm}|p{0.9cm}|p{0.9cm}|p{0.9cm}|p{0.9cm}|}
 \hline
   & \multicolumn{4}{|c}{\textbf{Without Balancing}} &  \multicolumn{4}{|c|}{\textbf{With level-I Balancing}} &  \multicolumn{4}{|c|}{\textbf{With level-II Balancing}}  \\
 \hline
  \textbf{Image \#}  &  \textbf{Se}  & \textbf{Sp} & \textbf{ACC} & \textbf{AUC} & \textbf{Se} & \textbf{Sp} & \textbf{ACC} & \textbf{AUC} & \textbf{Se} & \textbf{Sp} & \textbf{ACC} & \textbf{AUC} \\
 \hline
  1  &  45.39  &  99.63  &  95.18  &  94.62  &  76.40  &  97.00 & 95.32 & 96.32 & 80.99 & 97.58 & 96.22 & 98.05  \\
 \hline
 2  &  38.10  &  99.74  &  94.44  &  94.10  &  79.02  &  97.08  &  95.52  & 97.30  & 82.18 & 97.83 & 96.48 & 98.52   \\
 \hline
 
 Average (All Images)  & 40.07 & 99.68 & 94.47 & 93.96 & 78.5 & 96.57 & 94.99 & 96.75 & 81.57 & 97.65 & 96.22 & 98.23  \\
 \hline
  
\end{tabular}
\label{table 1}
\end{table*}

\section*{Datasets and Experimental Setup}
The Digital Retinal Images for Vessel Extraction (DRIVE) \cite{41} dataset is used for retinal segmentation. This dataset was collected in a study conducted in the Netherlands. $400$ individuals were recorded, where ages ranged between $25-90$. This dataset consists of randomly selected $40$ samples from the studied population. Retinal images for analysis were taken with a 3CCD camera that captured spherical areas from the center with a radius of $540$ pixels. Most samples $(N = 33)$ are medically normal, while there are signs of diabetic retinopathy in $7$ samples. These $40$ labeled color eye fundus images are equally distributed into train and test subcategories. For the training dataset, manual segmentation of each retinal fundus image's retinal blood vascular tree is available. At the same time, two expert segmentation ground truths are available for the test dataset. It is a usual practice to set one of them as the gold standard, whereas the second is used as a benchmark.

\begin{figure*}[ht!]
       \centering
       \captionsetup{justification=centering}
       \subfigure[Original image]{
       \includegraphics[width=.22\textwidth]{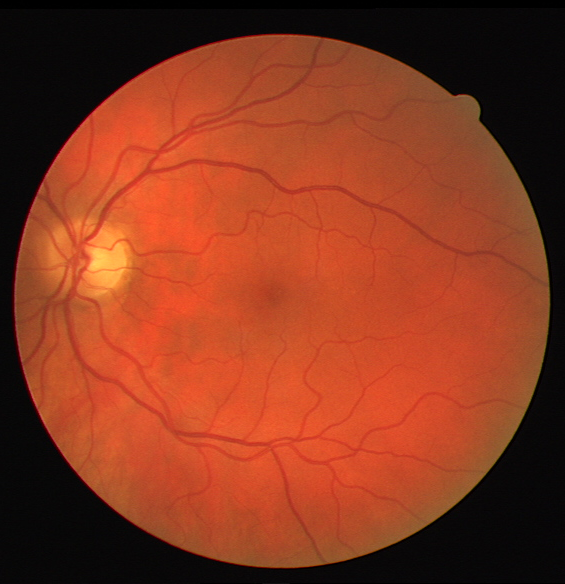}
        }
       \subfigure[Ground truth]{
       \includegraphics[width=.22\textwidth]{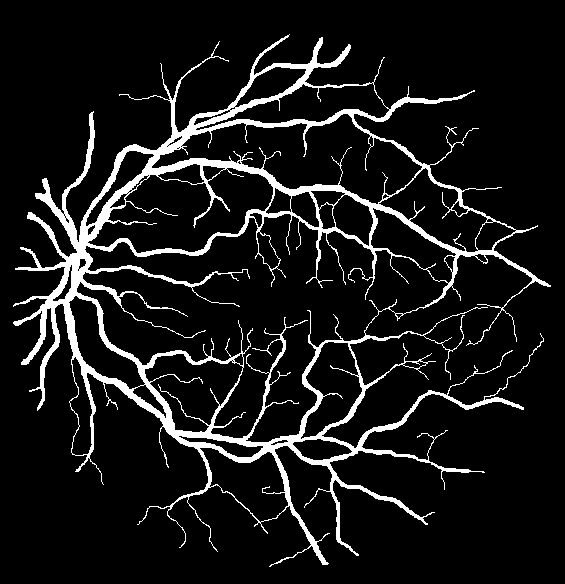}
       }
       \subfigure[Binary mask]{
       \includegraphics[width=.22\textwidth]{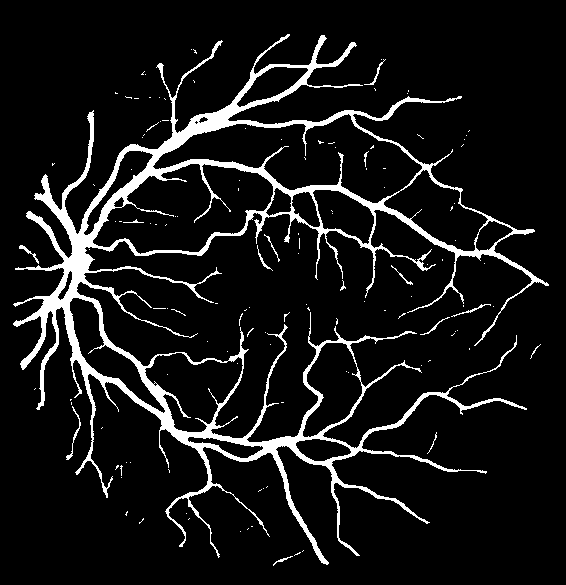}
       }
      \label{Fig 7 Row 1}
\end{figure*}

\begin{figure*}[ht!]
       \centering
       \captionsetup{justification=centering}
       \subfigure[Original image]{
       \includegraphics[width=.22\textwidth]{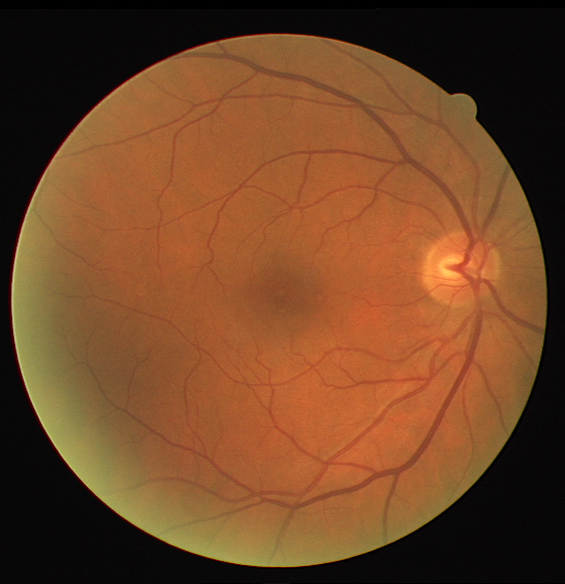}
        }
       \subfigure[Ground truth]{
       \includegraphics[width=.22\textwidth]{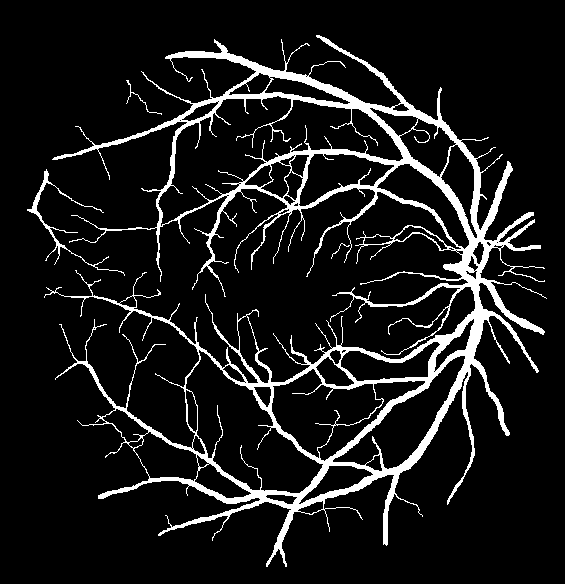}
       }
       \subfigure[Binary mask]{
       \includegraphics[width=.22\textwidth]{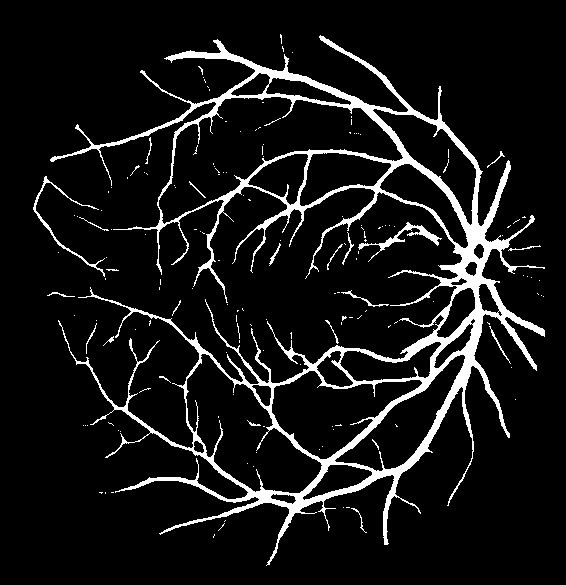}
       }
      \caption{Visual results for two sample DRIVE images.}
      \label{Fig 7 Row 2}
      \bigskip
\end{figure*}

\par Structured Analysis of the Retina (STARE) \cite{42} is fundus images database that contains $20$ retinal fundus images. The images are captured by a TopCon TRV-50 fundus camera. Among these images, $10$ images are pathology-free, while the remaining $10$ possess pathological abnormalities, which severely damage the anatomical structure of the eye. These abnormalities overlap blood vessels or sometimes make them completely complicated for analysis. This situation makes segmentation a much more complicated challenge to evaluate the performance in a more robust way. Like DRIVE, this dataset provides two sample sets of ground truth segmentation masks. The experiments in this paper were performed using Python. In particular, the Keras API with TensorFlow backend is used to construct the deep learning model. The proposed model is trained for $80$ epochs. All the experiments have been conducted on the Google Colab platform, which provides free shared cloud services and supports GPU-Tesla $K80, 2496$ CUDA cores, and $25$ GB GDDR5 VRAM.

\section*{Results and Discussions}
\par This section reports the quantitative and qualitative performance of the proposed system. We evaluate the impact of the class balancing algorithm in BLCB-CNN. Further, we report quantitative measures such as accuracy (ACC), sensitivity (Se), specificity (Sp), and area under the ROC curve (AUC) for DRIVE and STARE images. We also present the visual results of blood vessel segmentation for qualitative comparison with the corresponding ground truths. Finally, we compare the performance of the proposed model with various existing state-of-the-art methods in contemporary literature.
\begin{figure*}[t!]
       \centering
       \captionsetup{justification=centering}
       \subfigure[Original image]{
       \includegraphics[width=.22\textwidth]{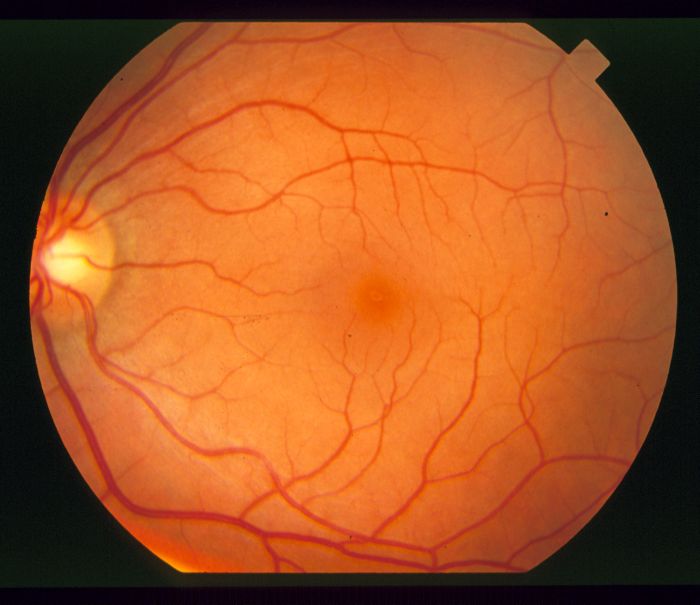}
        }
       \subfigure[Ground truth]{
       \includegraphics[width=.22\textwidth]{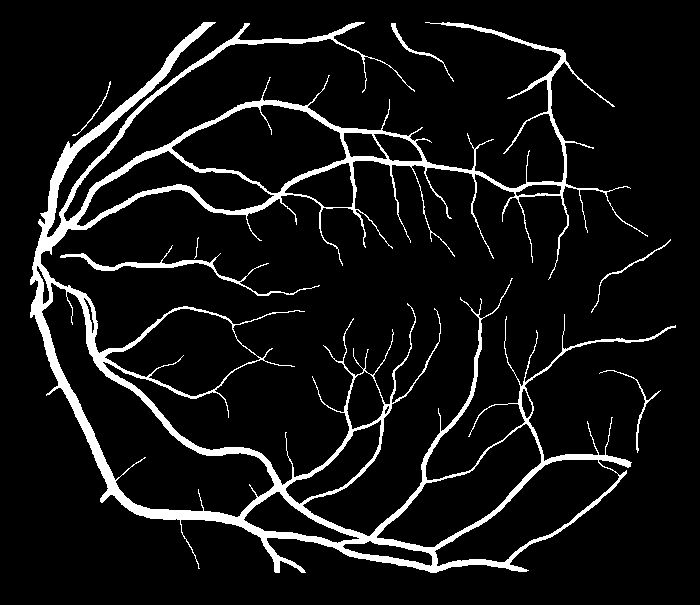}
       }
       \subfigure[Binary mask]{
       \includegraphics[width=.22\textwidth]{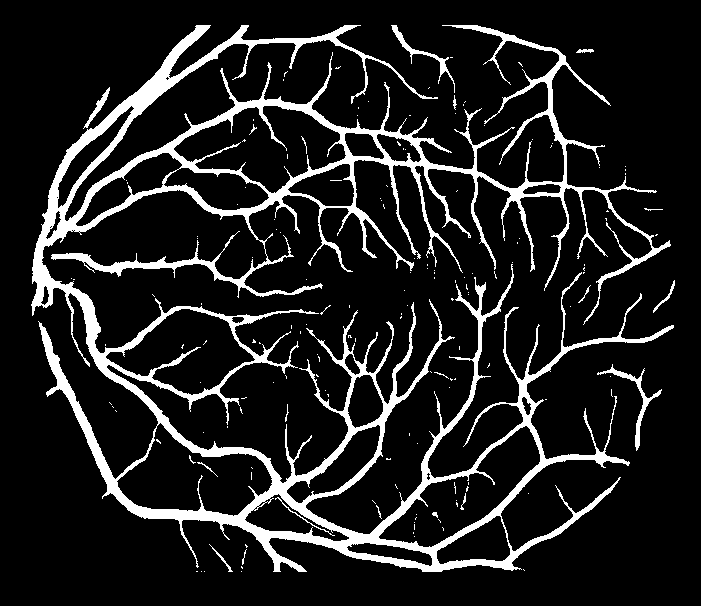}
       }
      \label{Fig 8 Row 1}
\end{figure*}

\begin{figure*}[t!]
       \centering
       \captionsetup{justification=centering}
       \subfigure[Original image]{
       \includegraphics[width=.22\textwidth]{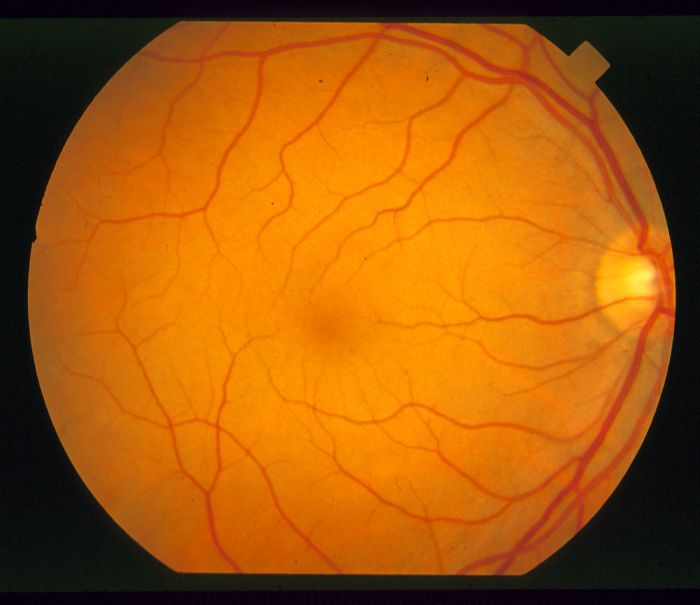}
        }
       \subfigure[Ground truth]{
       \includegraphics[width=.22\textwidth]{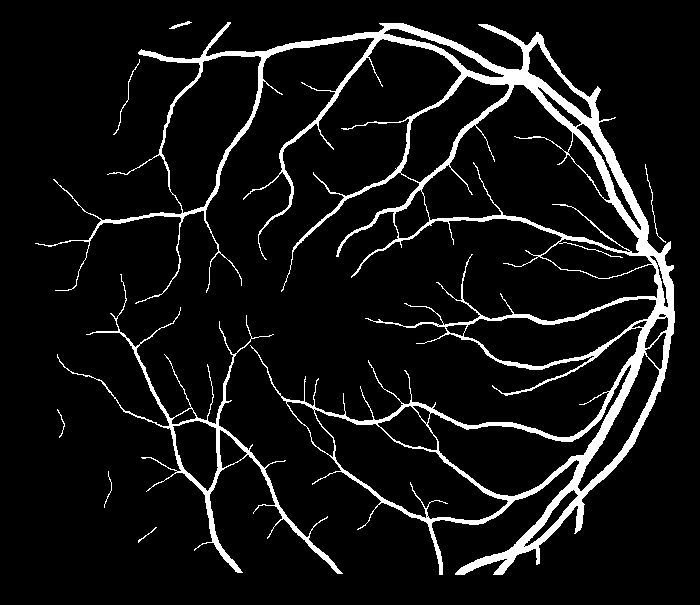}
       }
       \subfigure[Binary mask]{
       \includegraphics[width=.22\textwidth]{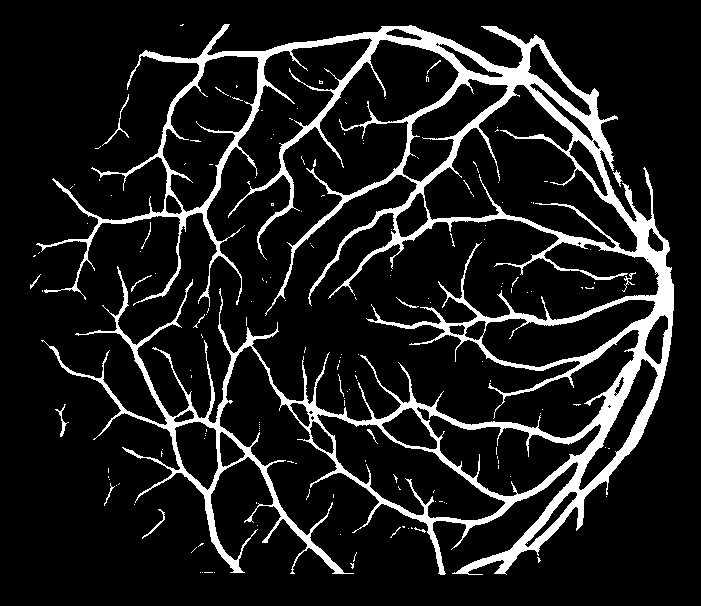}
       }
      \caption{Visual results for two sample STARE images.}
      \label{Fig 8 Row 2}
\end{figure*}
\subsection*{Ablation Studies}
\par We conducted extensive experiments to evaluate the impact of class balancing algorithms on the performance of the proposed model. The proceeding subsections explain the impact of different balancing components.\\
\textbf{Impact Analysis of Class Balancing:} The bi-level class balancing is an important part of the proposed system. This section investigates the impact of the two levels of class balancing separately under optimal hyper-parameters. First, the performance of the system is investigated with/without the Level-I balancing. Next, Level-II balancing is applied on top of Level-I balancing to investigate the exclusive impact of Level-II balancing. The following two sub-sections report both quantitative and qualitative results for the said experiments.\\
\textbf{Impact of Level-I Balancing:} Table I illustrates the performance of the CNN before and after applying Level-I and Level-II balancing on a couple of randomly sampled images from the DRIVE dataset. It should be noted that the proposed CNN is quite naïve in the absence of Level-I balancing (as evidenced by quite low sensitivity value). CNN favors non-vessel pixels, as it is the dominant class. However, Level-I balancing improves the sensitivity result, though specificity is a bit compromised now. However, as we will see in the next sub-section, this compromise is also normalized after the application of Level-II balancing. Nevertheless, we conjecture here that the proposed balancing algorithm augments even a simple CNN model (only ten layers) to produce state-of-the-art results. Figure 5 shows the visual results of segmenting the two sample images along with their ground truths. It can be observed in the visual analysis that the overall distinction of vessel pixels is much better after the application of Level-I balancing. However, the model is lacking in segmentation of thin vessels, even after the application of Level-I balancing. We conjecture this is due to the imbalanced distribution of thin and thick vessel pixels, which is overcome by Level-II balancing, as discussed in the next subsection. \\ 
\textbf{Impact of Level-II Balancing:} This section analyzes the impact of applying Level-II balancing on top of Level-I balancing to the same sample images as in the previous sub-section. Specifically, the outcomes of applying Level-I and Level-II balancing are compared and graphically demonstrated in Figure 6. The application of Level-II balancing results in a significant performance boost over Level-I balancing, as indicated by higher values of all performance measures. Similarly, Figure 6 shows the visual outcomes of Level-II balancing for the sample DRIVE images along with their ground truths. The output segmentation masks show the detection of thin retinal blood vessels much closer to the ground truth binary mask than the images for Level-I balancing. The red rectangle is highlighted as a focus area for conveniently comparing the results. It can be concluded from comparing the results that Level-II balancing detects most of the vessel’s pixels, especially thin vessels, thereby significantly improving the sensitivity. \\
\textbf{Results for DRIVE Test Images:} Quantitative results of BLCB-CNN for all DRIVE images are computed, which shows that the proposed method achieves an average accuracy of 96.22. The average sensitivity/specificity value of 81.57/97.65 also indicates the model’s ability to accurately segment vessel pixels (including thin vessels). The AUC results are also consistent with the accuracy obtained for all the images. Figure 7 demonstrates the visual outcomes for two sample DRIVE images. It can be observed that the proposed method segmented both thin and thick vessels much closer to ground truth binary masks. This is further evident from the focused rectangular areas of the output/ground-truth images.

\begin{table*}[t!]
\centering
\caption{Performance comparison with state-of-the-art techniques using DRIVE dataset.}
\begin{tabular}{|p{2.5cm}|p{6.6cm}|p{2.0cm}|p{0.8cm}|p{0.8cm}|p{0.8cm}|p{0.8cm}|}
\hline
\textbf{Ref.} & \textbf{Methodology} & \textbf{Validation Technique} & \textbf{Se} & \textbf{Sp} & \textbf{Acc} & \textbf{AUC} \\ \hline

Yan et al. \cite{6} & Three-stage deep learning model with ThickSegmenter, ThinSegmenter, and FusionSegmenter to separately segment thick and thin vessels and then fuse the results & Train-test split & 76.31 & 98.20 & 95.38 & 97.50 \\  \hline

Liskowski et al. \cite{23} & Deep Convolutional Neural Network with ReLU, Dropout, and Structured Prediction variants & Fixed split & 78.56 & 97.30 & 95.33 & 97.38  \\ \hline

Orlando et al. \cite{24} & Fully Connected Conditional Random Field (FC-CRF) trained with Structured Output SVM (SOSVM) using unary and pairwise features & Train-test split & 78.97 & 96.84 &-- &--  \\ \hline

Dasgupta et al. \cite{27} & Fully Convolutional neural network for structured prediction using cross-entropy loss and multi-label inference & Separate test data & 76.91 & 98.01 & 95.33 & 97.44 \\ \hline

Hu et al. \cite{29} & Multiscale convolutional neural network combined with fully connected conditional random fields and an improved cross-entropy loss function & Separate test data & 77.72 & 97.93 & 95.33 & 97.59 \\ \hline

Araujo et al. \cite{43} & Single-resolution fully convolutional network using raw fundus images with focal loss & Train-test split & 80.3 & 97.9 & 95.6 & 98.2 \\ \hline

Ma et al. \cite{New-1} & Modified U-Net architecture with decoder fusion module, context squeeze-and-excitation, and supervised fusion module incorporating attention mechanisms & Train-test split & 82.43 & 97.58 & 95.65 & 98.07 \\ \hline

Li et al. \cite{New-2} & A U-Net-based model with adaptive deformable convolution and adaptive parallel attention blocks & Train-test split & 82.39 & 97.69 & 95.74 & 98.18 \\ \hline

ResNet \cite{ResNet} & ResNet model for classifying each pixel into vessel and non vessel categories & Separate test data & 81.78 & 96.73 & 95.41 & 97.93 \\ \hline

\textbf{Proposed} & A deep CNN integrated with Bi-Level Class Balancing algorithm (Level-I for vessel/non-vessel balancing and Level-II for thick/thin vessel balancing) & Separate test data & \textbf{81.57} & \textbf{97.65} & \textbf{96.22} & \textbf{98.23} \\ \hline

\end{tabular}
\label{table 2}
\end{table*}

\subsection*{Performance Comparison with State-of-the-Art Approaches}
In this Section, we compares the performance of the proposed BLCB-CNN model with several existing methods. The proposed model is trained using 20 training images of the DRIVE dataset and evaluated on 20 test images of the same dataset. The comparison of quantitative performance measures on DRIVE dataset with state-of-the-art methods are presented in Table 2. The proposed BLCB-CNN model significantly advances retinal vessel segmentation performance by addressing one of the most persistent challenges in medical image analysis of data imbalance. Unlike many other models, such as three-stage deep learning model \cite{6} or adaptive U-Net \cite{New-2}, which primarily focus on architectural sophistication or multi-branch attention for feature extraction, the proposed BLCB-CNN uniquely integrates a BLCB technique. This balancing operates at two levels: vessel vs. non-vessel pixels and thick vs. thin vessel pixels. Our proposed BLCB-CNN model demonstrates a considerable improvements in sensitivity (81.57\%), an area where many competing models tend to falter due to their inability to accurately capture thin vessels which is a critical feature in early diagnosis of diabetic retinopathy and hypertensive retinopathy. Similarly, from a specificity point of view our model achieves a high value of 97.65\% and outperforming other methods like fully connected CRF model \cite{24} and structured prediction FCN model \cite{27}, which report specificity values of 96.84\% and 98.01\%, respectively. FCN-structured approach optimizes pixel-level spatial consistency, therefore it often overfit to thick vessels or background noise. However, proposed BLCB-CNN model on the other hand avoids overfitting through a carefully tuned dropout mechanism and balanced training dataset that improves generalization without the need for excessively deep or computationally expensive networks.

In terms of overall accuracy (96.22\%), the proposed BLCB-CNN outperform multiscale CNN model \cite{29} and modified U-Net with attention model \cite{New-1}, which typically range between 95.33\% and 95.74\%. This accuracy enhancement is attributed not only to the class balancing mechanism but also to the preprocessing pipeline (CLAHE + gamma correction) and post-processing (morphological erosion), that refine the vessel boundaries and remove spurious noise aspects often underutilized in other models. Most notably, the proposed BLCB-CNN delivers the highest AUC value (98.23\%) among all compared methods. The AUC is a robust indicator of the model's ability to distinguish between classes across different thresholds. Competing methods, such as the single-resolution FCN \cite{43} and adaptive U-Net model \cite{New-2} reach 98.20\% and 98.18\%, respectively. The BLCB-CNN's higher AUC signifies a better trade-off between sensitivity and specificity, particularly valuable in clinical settings where missing a vessel pixel (false negative) could be more detrimental than a false positive.

\subsection*{Performance Comparison with ResNet}
The proposed novel BLCB algorithm significantly addresses class imbalance in retinal vessel segmentation from fundus images. Its independence from model architecture ensures versatility and robustness. To confirm its effectiveness, the proposed BLCB algorithm is validated using ResNet50, following a similar pre-processing methodology for data preparation prior to training the neural network model. ResNet50 \cite{ResNet} is a robust variant of CNN within the residual networks family, designed explicitly to overcome the challenges of vanishing gradients during neural network training. This architecture is available in multiple depths, such as ResNet-18, ResNet-50 and ResNet-101. However, ResNet-50 stands out as a mid-sized variant that delivers outstanding depth and efficiency in image classification tasks. We combined our proposed BLCB algorithm with the ResNet50 model to validate BLCB as a general step in the processing pipeline that is independent of the actual prediction algorithm. Our experimental results demonstrate that this combination achieved good performance in terms of Se, Sp, Acc, and AUC with values of 81.78, 96.74, 95.41, and 97.94, respectively. These performance metrics are significantly high and comparable with those achieved by our proposed BLCB-CNN methodology. Thus, significant results across various networks using our novel BLCB algorithm showcase its generalizing capabilities that  establishes a convincing instance for other networks to adopt our algorithm to enhance the training processes and achieve promising results.


\subsection*{Cross Validation} Cross-validation is used to assess the performance of the proposed model on ten images from the STARE dataset. STARE images are more challenging due to the inherent lower quality and poor contrast of these images. However, the proposed model achieves reasonable performance on the STARE dataset images. Typically, training is conducted on STARE images using leave-one-out validation to test the model's validity. Very few works have provided cross-validation to evaluate the stability of the learned model on the STARE dataset. The average values 88.06/96.31 for Se/Sp are quite promising and demonstrate the significant performance of the proposed model. The average AUC value is 97.57\% that reflects the generalization of the proposed model on a completely unseen STARE dataset to demonstrate its generalization capability. Therefore, we, believe that our proposed methodology can be reliably applied for automatic retinal vessel segmentation of the clinical applications, such as a computer-aided diagnosis pipeline or automated vascular quantification or localization. Figure 8 depicts the visual outcomes for 2 sample images from the STARE dataset. As evident from a visual comparison of ground truth and output segmentation masks, thick vessels are detected quite accurately, especially near the fovea region. Similarly, thin vessel pixels at the branches are also segmented with reasonable visual match.

\section*{Conclusion and Future Work}
This paper presented a deep CNN-based fully automated vessel segmentation approach called BLCB-CNN to handle the primary challenge of segmenting thin blood vessels in retinal fundus images. The BLCB-CNN pipeline used a novel approach to combine CNN architecture with a Bi-Level class imbalance handling algorithm for overcoming the inherent inter- and intra-class imbalanced distribution. Moreover, we applied simple but effective pre- and post-processing techniques to obtain a more accurate output vessel tree. The proposed BLCB-CNN is trained on the training set of the DRIVE dataset and validated on the test set of the DRIVE dataset. The BLCB-CNN model produces promising results and sets a new benchmark for sensitivity, accuracy, and AUC measures on the DRIVE dataset compared to other state-of-the-art techniques. The influence of bi-level class balancing has also been studied, and it has been shown to impact the model performance positively. In the end, the proposed model's performance is validated on completely unseen images from STARE and DRIVE datasets to demonstrate its generalization capability. This dataset is produced from different fundus cameras and in clinical settings different from the training set. The results in this case, too, are encouraging, indicating the ability to accurately detect both the thick and thin vessel pixels. Given these findings, the proposed model can be a valuable resource for further research in surgical AI as well as a suitable candidate for transfer to clinical settings. This research work has multiple future directions. First, deep network models are always computational and data-hungry. The limited data for training may be a source of low performance, which can be overcome by various data augmentation techniques. In the future, applying a combination of data augmentation techniques is expected to boost performance in surgical AI. Second, the class balancing algorithm proposed in this research is based on an empirical study. In the future, information theory and machine learning-based class imbalance handling techniques (i.e., minimum redundancy, maximum relevance, mRMR) can be explored. By balancing the data, the model can converge faster, reducing training time and enabling faster deployment.


\section*{Declarations}
\section*{Ethics approval and consent to participate} 
Not applicable. 
\section*{Consent for publication} 
Not applicable. 
\section*{Data Availability} 
Data will be provided on request by first author Atifa Kalsoom.
\section*{Competing Interest}
The authors declare that they have no competing interests. 
\section*{Authors contributions} 
A. Kalsoom, M.A. Iftikhat, and A. Ali formulated the idea and designed the research; A. Kalsoom, A. Ali, and M.A. Iftikhat, performed the simulations; M.A. Iftikhat, H. Ali, Z. Shah, and S. Balakrishan analyzed the results; M.A. Iftikhat, A. Ali, H. Ali, Z. Shah, and S. Balakrishan supervised the work; A. Kalsoom, M.A. Iftikhat, and A. Ali wrote the original manuscript, H. Ali, Z. Shah, and S. Balakrishan revised the original and revised manuscript. All authors have read and agreed to this version of the manuscript. 

\section*{Acknowledgements}
Research reported in this publication was supported by the Qatar Research Development and Innovation Council (QRDI) grant number ARG01-0522-230266. Disclaimer: The content is solely the responsibility of the authors and does not necessarily represent the official views of Qatar Research Development and Innovation Council. 

This document is a preprint version of the manuscript. The final definitive article will be published by Springer Nature and is subject to changes and full copyright under their terms. 

\end{document}